\documentclass[reqno,12pt]{article}
\usepackage{amsmath}
\usepackage{bm}
\usepackage{a4wide,amssymb}
\usepackage{graphicx,xcolor}
\usepackage{epstopdf}
\usepackage{bbm}
\usepackage{mathrsfs}

\usepackage[titletoc,toc,title]{appendix}

\usepackage[refpage]{nomencl}
\usepackage{etoolbox}

 \newcommand{\nn}{\nonumber}
\newcommand{\R}{{\mathbb R}}

\newcommand{\g}{\gamma}\renewcommand{\d}{\delta}
           
\newcommand\om{\omega}

\newcommand{\e}{\varepsilon}

\def\s{\sigma}

\def\L{{\Lambda}}
\def\II{{\mathcal I}}

\def\RR{{\cal R}}
\def\II{{\cal I}}

\def\AA{{\cal A}}

\def\SS{{\cal S}}

\def\be{\begin{equation}}
\def\ee{\end{equation}}
\def\bea{\begin{eqnarray}}
\def\eea{\end{eqnarray}}
\def\ni{\noindent}
\def\nn{\nonumber}

\def\d{\delta}
\def\o{\omega}
\def\b{\beta}

\usepackage{multicol}
\makenomenclature
\usepackage{xpatch}
\makeatletter
\xapptocmd\thenomenclature{\let\@item\nomencl@item\def\nomencl@width{0pt}}{}{}
\let\nomencl@item\@item
\xpretocmd\nomencl@item{\nomencl@measure{#1}}{}{}
\def\nomencl@measure#1{%
  \sbox0{#1}%
  \ifdim\wd0>\nomencl@width\relax
    \edef\nomencl@width{\the\wd0}%
  \fi
}

\xpatchcmd\thenomenclature
  {\section*{\nomname}}
  {\begin{multicols}{2}[\section*{\nomname}]}
  {}{}
\xpatchcmd\thenomenclature
  {\chapter*{\nomname}}
  {\begin{multicols}{2}[\chapter*{\nomname}]}
  {}{}

\xapptocmd\endthenomenclature{%
  \immediate\write\@mainaux{\global\nomlabelwidth\nomencl@width\relax}%
  \end{multicols}
}{}{}
\makeatother

\begin{document}

\begin{center} 
{\bf \LARGE{A kinetic model for epidemic spread}}

\vspace{1.5cm}
{\large M. Pulvirenti$^{1}$ and S. Simonella$^{2}$}

\vspace{0.5cm}
{$1.$ \scshape {\small Dipartimento di Matematica, Universit\`a di Roma La Sapienza\\ 
Piazzale Aldo Moro 5, 00185 Rome -- Italy,  and \\
 International Research Center M\&MOCS, Universit\`a dell'Aquila, \\ Palazzo Caetani, 04012 Cisterna di Latina -- Italy.}  
 \smallskip

$2.$ {\small UMPA UMR 5669 CNRS, ENS de Lyon \\ 46 all\'{e}e d'Italie,
69364 Lyon Cedex 07 -- France}}

\end{center}

\vspace{1.5cm}

\noindent
{\bf Abstract.} We present a Boltzmann equation for mixtures of three species of particles reducing
to the Kermack-McKendrick (SIR) equations for the time-evolution of the density of infected agents in an isolated population. The kinetic model is potentially more detailed and might provide information on space mixing of the agents.

\vspace{0.5cm}\noindent
{\bf Keywords.} Boltzmann equation; SIR model; low-density limit; stochastic particle system; forward cluster.

\vspace{1.5cm}

\section{Boltzmann--SIR equations} \label{sec:1}
\setcounter{equation}{0}    
\def\theequation{1.\arabic{equation}}

Consider a population of identical individuals (particles) moving in the physical space and interacting upon contact. One (or several) of the individuals, say particle $1$, has an infected status  at time zero. As the dynamics runs, the infection can be transmitted, at the interaction times, to the individuals entering in contact with $1$ or with the newly infected individuals. A cluster $\{i_1,i_2,\cdots\}$ of infection grows in time, determined by the particle evolution: an individual is potentially infected at time $t>0$ if it is involved, directly or indirectly, in the forward-in-time dynamics of $1$. The ``forward cluster of particle $1$'' (according to a terminology of \cite{Aoki,PS20}) is represented symbolically in the picture below. 
\begin{figure}[htbp] 
\centering
\includegraphics[width=3in]{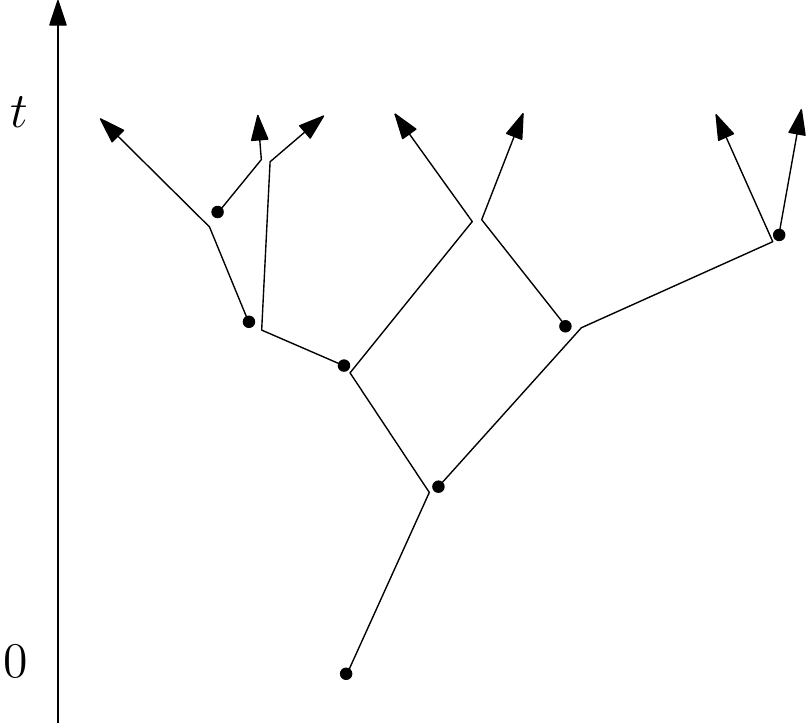} 
\end{figure}

For concreteness, we may want to fix an idealized mechanical setting. Let us then proceed, as it is customary in kinetic theory, by looking at $N$ hard spheres of unit mass and diameter $\e>0$. The balls move in $\L \subset \R^d$, $d=2,3$, and interact through elastic collisions.
Each particle flies freely with constant velocity and, when two hard spheres collide with positions $x, x_*$ at distance $\varepsilon$ and incoming velocities $v, v_*$ 
the latter are instantaneously transformed to outgoing velocities $v', v'_*$ by  
the relations 
\be
\begin{cases}
\displaystyle v'=v-\om [\om\cdot(v-v_*)] \\
\displaystyle  v'_*=v_*+\om[\om\cdot(v-v_*)]
\end{cases}\;,\label{eq:coll}
\ee
where $\o$ is the normalized relative distance 
$\o = (x-x_*)/|x-x_*|=(x-x_*)/\varepsilon \in {\mathbb S}^{d-1}$.

We shall mimic the basic model in the mathematical theory of epidemics \cite{KMK}, by means of several assumptions.
There are three different species of particles:
$S$, $I$ and $R$ which stay for susceptible, infected and recovered, respectively. Upon collision between a particle of type $S$ and a particle of type $I$, the reaction
 $$
 S+I \to I+I
 $$
occurs instantaneously with rate $\b \in [0,1]$. All the other collisions do not change the particle type but, in addition, a decay $$I\to R$$ occurs with rate $\g \in [0,1]$. Note that the population size is fixed (no deaths), and that the infection implies complete immunity. Finally for simplicity, we shall assume that $\b$ and $\g$ are constants (they do not depend on time).

We are relying on the idea that the details of the interactions should not be of crucial importance (see \cite{St20} for a recent popular article simulating a similar system of particles). The main features are instead the following.
\begin{itemize}
\item The interactions are binary, and localized.
\item The number of interactions per unit time is expected to be finite.
\item The qualitative behaviour is independent of the number of particles $N$, provided that this is large in a suitable scaling limit.
\item A statistical description is appropriate.
\end{itemize} 

\ni Under these assumptions, the Boltzmann equation for rarefied gases provides a tool of investigation. 

Let us perform the so called Boltzmann-Grad limit \cite{Gr49} on the hard sphere system under consideration. Denoting the one-particle distribution functions by
 \bea
& f_S=f_S(t,x,v)\nn\\ 
& f_I=f_I(t,x,v)\nn\\
& f_ R=f_R(t,x,v)\nn
\eea 
for the three species of particles, we obtain the following set of equations:
\begin{equation}
\begin{cases}
  \left(\partial_t + v \cdot \nabla_x\right) f_S= Q(f_S,f_S) + Q(f_S,f_R)+(1-\b)Q(f_S,f_I)-\b Q_-(f_S,f_I)  \\
    \left(\partial_t + v \cdot \nabla_x\right) f_I = Q(f_I,f) +\b Q_+(f_S,f_I)-\g f_I\\
     \left(\partial_t + v \cdot \nabla_x\right) f_R = Q(f_R,f)+\g f_I\\
\end{cases}\;\;, \label{eq:SIR_Boltzmann}
\end{equation}
where 
$$f=f_S+f_I+f_R$$
and $Q $ is Boltzmann's operator (expressed in asymmetric form)
\bea
&& Q= Q_+ - Q_- \nn\\
&& Q_+(f,g)(v) := \int_{\R^d} \int_{ {\mathbb S}^{d-1}} B(\o; v-v_*)f(v')g(v'_*) \, d\o \, dv_* \;\;\;.\nn\\
&& Q_-(f,g)(v) := f(v) \int_{\R^d} \int_{ {\mathbb S}^{d-1}} B(\o; v-v_*)g(v_*) \,  d\o \, dv_* \nn 
\eea
Note that the sum $f=f_S+f_I+f_R$ satisfies the classical Boltzmann equation
$$ 
\left(\partial_t + v \cdot \nabla_x\right) f = Q(f,f)\;.
$$
Here we chose
$
B(\o; v-v_*)=\left(\o \cdot (v-v_*)\right)\,\mathbbm{1}\left(\o \cdot (v-v_*)\geq 0\right)\;,
$
corresponding to the hard-sphere cross section. However as said above, conclusions drawn from 
the kinetic model should not be very sensitive on the interaction rule;
e.g.\,we shall consider as well different kernels $B(\o; v-v_*) \geq 0$ such that
$$\
\int_{ {\mathbb S}^{d-1}} B(\o; v-v_*)  d\o = |v-v_*|^{b}\;,
$$
for some $b \geq 0$.

In the second part of this exposition we will give more details on the passage from the particle dynamics to \eqref{eq:SIR_Boltzmann}.
Before that, we make a few elementary remarks on the equations themselves.

\subsection {Maxwell collisions: Kermack-McKendrick equations}

Averaging \eqref{eq:SIR_Boltzmann} over velocities, the $Q$ operators vanish (because $\int Q_+=\int Q_-$) and, in the spatially homogeneous case (no dependence on $x$),  the expected fractions of individuals of the species $A \in \{ S,I,R \}$, $A(t)=\int f_A(t,v) dv$, satisfy the equations
\be \label{eq:SIR'}
\begin{cases}
  \dot S =  -\b \int \, |v-v_*|^b \,f_S(v)\, f_I(v_*) \,dv\, dv_*\nn\\
    \dot I =\b \int \, |v-v_*|^b \,f_S(v)\, f_I(v_*) \,dv\, dv_* -\g I \nn\\
     \dot R = \g I \nn\\
\end{cases}\;.
\ee
These equations are not closed, except when dealing with ``Maxwellian molecules'' (case $b=0$ \cite{Bo88}) for which we get
\be \label{eq:SIR'}
\begin{cases}
  \dot S = -\b IS  \\
    \dot I = \b SI -\g I \\
     \dot R = \g I \\
\end{cases}\;,
\ee
namely the epidemiology model of \cite{KMK} in the case of time-independent rates.
This model has been analysed and used extensively, and several generalizations have been conceived; see e.g.\,\cite{AM79,BC-C01,Mu02,HLM}, and the many references therein.
The kinetic equation \eqref{eq:SIR_Boltzmann} stands as an extension accounting for dependence on space and velocity of the individuals.

To remind the reader of the original motivations for such SIR models (\cite{Ro16,KMK}), we recall that an epidemic is not necessarily terminated by the exhaustion of the susceptible individuals, nor by the extinction of the virulence. This is apparent from \eqref{eq:SIR'}, over a threshold value of the density. 
Setting indeed $A_\infty := \lim_{t\to \infty} A(t)$, $A_0 = A(0)$ and 
$
R(t)=R_0 + \g \int_0^t I(\tau) d\tau 
$
(showing that $I(t) \to 0$ as $t \to \infty$), one has that $
\frac {d S}{dR}= - \frac \b \g S
$
and hence (by $R_\infty+S_\infty=1$ and the assumption $R_0 = 0$)
$
S_\infty= S_0\, e^{- \frac \b \g\left(1-S_\infty\right) } \;,
$ or
\be
e^{ -\frac \b \g S_\infty} \,\frac \b \g S_\infty= S_0\, \frac \b \g e^{- \frac \b \g }\;.
\label{eq:Sinf}
\ee
Since $\max y \,e^{-y}=\frac 1 e$, given a value of $\b/\g$ one can find non vanishing solutions for $S_\infty$, provided that $I_0 = 1-S_0$ is sufficiently large.

\subsection{Confinement}

The model can be easily adapted to investigate several different situations. Examples might be boundary conditions or external potentials, imposing internal spatial constraints or local enhancing of density. There has been recent intense interest in the effects of isolation of individuals, and of the reduction of social mixing, by means of physical distancing measures (\cite{LGW20, PLR20}). At the level of \eqref{eq:SIR_Boltzmann}, the energy can be used as simple parameter regulating the interaction rate. 

We just make an example of one adaptation of  \eqref{eq:SIR_Boltzmann}, intended to model a confinement effect.
Following \cite{St20} we assume that, for each species, there are two types of particles: wandering and confined.
We denote by $g_A$, $A\in \{S,I,R\}$, the distribution of confined particles, while we maintain the notation $f_A$ for the wandering particles. The distribution of the species $A$ is $h_A:=f_A+g_A$ and $f=\sum_A h_A$.
 Wandering particles have mass $m_w=1$, while confined particles have mass $m_c =+ \infty$ and zero velocity. 
The distribution  $g_A$ is proportional to a Dirac delta in velocity. Confined particles are frozen and their total distribution is stationary: 
$$
g_S(t,x)+g_I(t,x)+g_R(t,x)=\text {const.} \quad \forall t\;.
$$ 
The collision law becomes
\be
\begin{cases}
 v'=v-\frac{2m_*}{m+m_*}\,\o [\om\cdot(v-v_*)] \nn\\
 v'_*=v_*+\frac{2m}{m+m_*}\,\o [\om\cdot(v-v_*)] \nn
\end{cases}
\ee
where $m, m_*$ are the masses of the incoming particles,
and Eq.\,\eqref{eq:SIR_Boltzmann} is replaced by
\begin{equation}
\begin{cases}
 \left(\partial_t + v \cdot \nabla_x\right) f_S= Q(f_S,h_S) + Q(f_S,h_R)+(1-\b)Q(f_S,h_I)-\b Q_-(f_S,h_I)  \\
    \left(\partial_t + v \cdot \nabla_x\right) f_I = Q(f_I,f) +\b Q_+(f_S,h_I)-\g f_I\\
     \left(\partial_t + v \cdot \nabla_x\right) f_R = Q(f_R,f)+\g f_I.\\
     \dot{g_S} = - \b Q_-(g_S,f_I)\\
     \dot{g_I} =\b Q_+(g_S,f_I)-\g g_I\\
     \dot{g_R}=\g g_I
\end{cases}\;. \label{eq:SIR_Boltzmann1}
\end{equation}
In the spatially homogeneous case, integrating Eq.s \eqref{eq:SIR_Boltzmann1} in $v$, calling $A_w = \int f_A\, dv$ and $A_c=\int g_A\,dv$, $A=S,I,R$, we obtain
\begin{equation}
\begin{cases}
  \dot{{S}}_w = -\b \int  |v-v_*|^b \, f_S(v)\, h_I(v_*)\,dv\, dv_*\,  \\
   \dot{{I}}_w = \b \int|v-v_*|^b \, f_S(v)\, h_I(v_*) \,dv\, dv_*\, -\g  {I}_w \\
     \dot{{R}}_w = \g {I}_w \\
    \dot{ {S_c}}=  -S_c\int  |v_*|^b \, f_I(v_*)\,dv_*\,\\
     \dot{ {I_c}}=  S_c\int |v_*|^b \, f_I(v_*) \,dv_*\, - \g I_c\\
    \dot{{R_c}} =\g I_c
\end{cases}\;. \nn\label{eq:SIRg}
\end{equation}
Again, Eq.s\,\eqref{eq:SIRg} reduce to a standard SIR model in the case of Maxwellian molecules:
\begin{equation}
\begin{cases}
  \dot{{S}}_w = -\b   {{S}}_w\left( {{I}}_w+I_c\right) \\ 
   \dot{{I}}_w = \b   {{S}}_w\left( {{I}}_w+I_c\right)-\g  {I}_w \\
     \dot{{R}}_w = \g {I}_w \\
    \dot{{S_c}}=  -{S_c}  {{I}}_w\\
     \dot{{I_c}}=  {S_c}  {{I}}_w - \g I_c\\
    \dot{{R_c}} =\g {I_c}.
\end{cases}\; . \label{eq:SIRg1}
\end{equation}

\subsection{Related problems}

The kinetic model presented above should be interpreted as a remark of mathematical physics character: we do not pretend that it can be of use in epidemiology. It is more detailed than the classical SIR, insofar it includes space and velocities of the agents. Presumably, its main potential interest in applications is the identification of spatial patterns having an impact on the history of epidemics.
Moreover, a dynamical representation in terms of forward (or backward) clusters would provide information on the tracing of the infection.
We comment next  on a few other problems arising naturally.

The typical question concerning SIR equations is determining the long-time behaviour in relation with the parameters $\b,\g$, and its dependence on local characteristics of the initial data. We are interested in masses, but also in local densities in the presence of spatial inhomogeneities. From the mathematical side, little can be done, but the problem is suited to numerical investigation. In analogy with gas dynamics, it is natural to use stochastic methods, as we will discuss in the next section.

At the theoretical level, it would be interesting to detect large scale limits and derive, starting from \eqref{eq:SIR_Boltzmann}, equations for locally conserved quantities. Eq.\,\eqref{eq:SIR_Boltzmann} can be useful in fact  for limited amounts of time. Preliminarily, one should characterize the equilibria.
Let $F_A = \lim_{t \to 0} f_A$ be the asymptotic distributions. Then we expect $F_I=0 $, and the other two distributions should satisfy
\begin{equation}
\begin{cases}
 Q(F_S,F_S) + Q(F_S,F_R)=0 \\
  Q(F_R,F_R)+Q(F_R,F_S)=0
\end{cases}\;.\nn
\end{equation}
The latter equation is satisfied if both $F_S$ and $F_R$ are Maxwellians
$$
F_A=A_\infty \,\, \frac {e^{- \frac {(v-u)^2} {2 \s^2 }}} { (2 \pi \s^2 )^{d/2}}
$$
for some constants $S_\infty$ and $R_\infty$, with $\s$ and $u$ determined by the initial conditions. $A_\infty$ would be obtained as in \eqref{eq:Sinf}. Notice that, when $f=f_S+f_I+f_R$ is a global equilibrium, a solution $(f_S,f_I,f_R)$ of Eq.\,\eqref{eq:SIR_Boltzmann} for $b=0$ is given by the same global equilibrium with densities $S(t), I(t), R(t)$ driven by 
\eqref{eq:SIR'}.

\section{Particle systems} \label{sec:2}
\setcounter{equation}{0}    
\def\theequation{2.\arabic{equation}}

\subsection{Stochastic particle system}

In this section we introduce a particle system yielding, in a suitable scaling limit, kinetic equations of type \eqref{eq:SIR_Boltzmann}. The interest of this dynamics is twofold. First, it can be considered as a microscopic model to be accepted as the phenomenology, covering a large variety of kernels $B$. 
It would be somewhat funny to believe that the laws of Newton can be used to describe efficiently the interaction among individuals. On the other hand, we do not know so much concerning the details of such interactions, thus a stochastic collision appears to be more robust than a deterministic one.
Secondly, the particle scheme corresponds numerically to the direct simulation Monte Carlo method, widely used to approximate rarefied gas dynamics. There are several variants of such methods (\cite{Bird,RW05}). Below, we will deal with an inhomogeneous Kac model for three species with reactions \cite{Kac}.

We start by regularizing the collision operator \eqref{eq:SIR_Boltzmann}. The strictly local interaction is smeared as follows:
\bea
&& Q^h= Q^h_+ - Q^h_- \nn\\
&& Q^h_+(f,g)(x,v) := \int_{\R^d}  \int_{\R^d} \int_{{\mathbb S}^{d-1}} B(\o; v-v_*) \, h(|x-y|)\,f(x,v')g(y,v'_*) \, d\o\, dv_* dy  \nn\\
&& Q_-^h(f,g)(x,v) :=  \int_{\R^d}  \int_{\R^d} \int_{{\mathbb S}^{d-1}} B(\o; v-v_*) \, h(|x-y|)\,f(x,v) g(y,v_*) \,  d\o\, dv_* dy  \nn
\eea
where $ h: \R^+ \to \R^+$
 is a smooth approximation of the delta function.

To simplify the notation we limit ourselves to the  case of \eqref{eq:SIR_Boltzmann} with $\b = 1$, being the more general cases a trivial extension. We therefore consider the following equations:
\begin{equation}
\begin{cases}
 ( \partial_t+v\cdot \nabla_x)   f_S = Q^h(f_S,f_S) + Q^h(f_S,f_R)- Q^h_-(f_S,f_I)  \\
  ( \partial_t+v\cdot \nabla_x) f_I = Q^h(f_I,f) + Q^h_+(f_S,f_I) -\g f_I\\
 ( \partial_t+v\cdot \nabla_x) f_R = Q^h(f_R,f)+\g f_I.
\end{cases}\;. \label{eq:SIR_Boltzmann_h}
\end{equation}
We can pass to the limit $Q^h \to Q$ inside \eqref{eq:SIR_Boltzmann_h},
whenever we have a smooth solution of the initial value problem.

 We shall indicate by $\AA = \SS, \II, \RR \subset \{1,2,\cdots,N\}$ the (random) disjoint sets of particles of type $A = S,I,R$ respectively. They form a partition of $\{1,2,\cdots,N\}$, so that the
 process  $Z_N: \R^+ \to {\cal X} $, $Z_N = Z_N(t) = (z_1(t),\cdots,z_N(t))$, $z_i = (x_i,v_i)$, takes values in
 $$
 {\cal X} =\bigcup_{ \SS,  \II, \RR} {\cal X} (  \SS,  \II, \RR)\,, \quad
 {\cal X} (  \SS,  \II, \RR)= \big\{ \left(Z_{\SS}, Z_{\II}, Z_{\RR} \right)\big\}\;
 $$
 with $$ |\SS| + |\II| + |\RR| = N\;,$$
and $z_i \in \L \times \R^d$.
Here $|\AA|$ denotes the cardinality of the set $\AA$. The configuration of particles in the three species are  $Z_{\SS} = \left( z_{s_1}, z_{s_2},\cdots\right)$, $Z_{\II} = \left( z_{i_1}, z_{i_2},\cdots\right)$ and $Z_{\RR} = \left( z_{r_1}, z_{r_2},\cdots\right)$, respectively.

Le us define the time evolution. Particles move freely for a random time, exponentially distributed with intensity scaling like $N$. Then two particles are randomly chosen, say particles $j$ and $k$, according to $\int B(\o; v_j-v_k)\,h(|x_j-x_k|) \,d\o$ and their velocities are updated as in \eqref{eq:coll} with $\o \sim B(\cdot\,; v_j-v_k)$. If the pair of colliding particles is of type $ (A,A )$ or $(S,R)$ or $(I,R) $, the particles do not change their species. If the pair is of type $(S,I)$, then the outgoing pair is of type $(I,I)$. 
We abbreviate from now on $h_{j,k}= h(|x_j-x_k|)$, and we denote by $J_{jk}$ the linear operator transforming  the velocities $j$ and $k$ to a postcollisional pair with scattering vector $\o$. The generator of the process reads
$$
 {\cal L}= {\cal L}_0+ {\cal L}_i +{\cal L}_d
$$
where $ {\cal L}_0=\sum v_i \cdot \nabla_{x_i}$ is the generator of the free  motion, 
\begin{eqnarray}
 {\cal L}_i\phi (Z_N) &&= \frac{1}{N} \sum_{j \in \SS} \sum_{k \in \II}\int B(\omega; v_j-v_k) h_{j,k} \nn\\
&&\ \ \ \ \ \ \ \ \ \ \times\left(
J_{jk}\phi\left(Z_{\SS \setminus \{j\}},
Z_{\II \cup \{j\}}, Z_{\RR}
\right) - \phi(Z_N) \right) d\o\nn\\
&& +\frac{1}{N} \left(\sum_{j \in \SS} \sum_{k \in \RR} + \sum_{j \in \II} \sum_{k \in \RR}\right)\int B(\omega; v_j-v_k)h_{j,k}
\left(
J_{jk}\phi\left(Z_N
\right) - \phi(Z_N) \right) d\o\nn\\
&& +\frac{1}{2N} \sum_{\AA = \SS,\II,\RR}\;\sum_{ \substack{j,k \in \AA \\ j \neq k}} \int B(\omega; v_j-v_k) h_{j,k}
\left(
J_{jk}\phi\left(Z_N
\right) - \phi(Z_N) \right) d\o\;, \label{eq:gener}
\end{eqnarray}
and
\be
\label{d}
 {\cal L}_d\phi ( Z_N)= \g \sum_{i \in \II}  
\left(\phi ( Z_{\SS}, Z_{\II \setminus \{i\}}  ,     Z_{\RR \cup \{i\} } ) - \phi ( Z_N)\right)\;.
\ee

We choose now test functions of the form
$$
\phi_A(Z_N) = \frac{1}{N} \sum_{\ell \in \AA} \varphi(z_\ell ) 
$$
and focus, for instance, on the case $\AA = \SS$. We have that ${\cal L}_d\phi_S = 0$.
Evaluating Eq.\,\eqref{eq:gener} in $\phi_S$ we notice that, given $j$ and $k$, 
all the terms with $\ell \neq j,k$ cancel out. In the second line of
\eqref{eq:gener} we find
$$
\sum_{\substack{\ell \in \SS \\ \ell 	\neq j} }J_{jk}\,\varphi(z_\ell)
- \sum_{\substack{\ell \in \SS } }\varphi(z_\ell) = - \varphi(z_j)\;.
$$
Therefore
\begin{eqnarray}
\label{final}
 {\cal L}_i\phi_S (Z_N) &&= - \frac{1}{N^2} \sum_{j \in \SS} \sum_{k \in \II} \int B(\omega; v_j-v_k) \,h_{j,k}\, 
 \varphi(z_j)  d\o\nn\\
&&\ +\frac{1}{N^2} \sum_{j \in \SS} \sum_{k \in \RR}\int B(\omega; v_j-v_k) h_{j,k} 
\left( \varphi(x_j,v'_j) - \varphi(z_j)  \right) d\o\nn\\
&&\ +\frac{1}{2N^2}\sum_{ \substack{j,k \in \SS \\ j \neq k}} \int B(\omega; v_j-v_k) h_{j,k} 
\left( \varphi(x_j,v'_j) + \varphi(x_k, v'_k) - \varphi(z_j) - \varphi(z_k) \right) d\o\;. \nn\\
\end{eqnarray}

Next, we introduce a probability measure with density
$
W^N : {\cal X}\to \R^+\;,
$
assumed to be symmetric in the exchange of the particle labels within each one of the species.
An example is provided by the fully factorized (chaotic) state, which we shall assume, to fix ideas, as initial distribution of the particle process:
$
W^N(0)= f_0 ^{\otimes N}\nn
$
with $f^0=\sum_A f_A^0,  A=(S,I,R)$, where $f_A^0$ are the initial data for \eqref{eq:SIR_Boltzmann_h}.
We further denote by $f^N_{A} = f^N_{A}(z)$ the one-particle marginals of $W^N$, defined as
$$
\int f^N_{A}(z) \varphi (z) dz= \int W^N(Z_N) \phi_A(Z_N) dZ_N\;.
$$
It is the probability density of finding a particle of type $A$ in $z$.
Similarly, $f^N_{A_1,A_2}= f^N_{A_1,A_2}(z_1,z_2) $ denotes the two-particle marginal, namely the probability density of finding two particles of type $A_1$ and $A_2$ in $z_1$ and $z_2$: 
$$
\int  f^N_{A_1,A_2} (z_1,z_2) \varphi(z_1,z_2)dz_1 dz_2 = \int W^N(Z_N) \phi_{A_1,A_2} (Z_N) dZ_N
$$
for $ \phi_{A_1,A_2} (Z_N)=\frac 1 {N(N-1)} \sum_{j\in {\cal A}_1} \sum_{\substack{k\in {\cal A}_2 \\ k \neq j}} \varphi (z_j,z_k)$.
Even though the initial measure is factorized, the time-evolved density $W^N(t)$ is not, due to correlations generated by the dynamics. The factorization is however recovered in the limit $N \to \infty$ and
\be
\label{pc}
f^N_{A_1,A_2} (z_1,z_2)  \approx f^N_{A_1} (z_1)  f^N_{A_2} (z_2) \;.
\ee 

We are ready to compute
$$
\frac d {dt} \int W^N(t)\, \phi_S =\int W^N(t) {\cal L}\phi_S\;.
$$
Using \eqref{final}, the definition of marginal and \eqref{pc}, we deduce that, as $N \to \infty$,
\be
\label{finaleq}
\frac d {dt} \int f^N_{S}(t)\varphi \approx \int f^N_S\, \left(v \cdot \nabla_x \varphi\right)+\int Q^h  (f^N_S, f^N_S)\varphi 
 +\int Q^h  (f^N_S, f^N_R) \varphi- \int Q^h_- (f^N_S, f^N_I)\varphi\;, \nn 
\ee
that is the first equation of \eqref{eq:SIR_Boltzmann_h} in weak formulation.

The other two equations can be recovered similarly. For $A=I$, Eq.\,\eqref{d} yields
$$
{\cal L}_d \phi_I ( Z_N )= \frac \g N \sum_{i \in \II}  
\left(\sum_{\ell \in \II \setminus {\{i\}}} \varphi (z_\ell) - \sum_{\ell \in \II } \varphi (z_\ell)\right) =- \frac \g N  \sum_{i \in \II}  \varphi (z_i)\;,
$$
while in the second line of
\eqref{eq:gener} we find 
$$
\sum_{\ell \in \II \cup \{j \} } J_{jk}\,\varphi(z_\ell)
- \sum_{\ell \in \II } \varphi(z_\ell) = J_{j,k} \varphi(z_j)+ \left(J_{j,k} \varphi(z_k) -\varphi(z_k) \right) 
$$
so that
\be
\label{finaleq}
\frac d {dt} \int f^N_{I}(t) \approx \int f^N_I\, \left(v \cdot \nabla_x \varphi\right)+\int Q^h  \left(f^N_I, \sum_A f^N_A\right)\varphi 
 +\int Q^h_+  (f^N_S, f^N_I) \varphi- \g\int f^N_I \varphi\;, \nn 
\ee
which is the second equation of \eqref{eq:SIR_Boltzmann_h}.

\subsection{Mechanical system}

We briefly come back to the deterministic particle model, which was our starting point. 
That is, $N$ hard spheres of diameter $\e$ moving in the physical space and colliding elastically, with 
reactions simulating infection and recovery. 
We call this system ``mechanical'' as the interaction is deterministic. Clealry there is still stochasticity in the reactions and, strictly speaking, we are dealing again with a stochastic process.

We can easily adapt to this case the formal arguments of the previous section. The process $Z_N$ takes still values in $\cal X$, but in addition the hard core exclusion is imposed $\min_{i \neq j} |x_i -x_j|>\e$.
In the generator \eqref{eq:gener}, $1/N$ is replaced by $\e^{d-1}$, $B$ is the hard-sphere kernel $\left(\o \cdot (v_j-v_k)\right)\,\mathbbm{1}\left(\o \cdot (v_j-v_k)\geq 0\right)$, $h_{j,k}$ is absent and the operator $\left(J_{j,k}-1\right)$ is replaced by $\left(\d (x_k-x_j -\o \e )J_{j,k} - \d (x_k-x_j +\o \e )\right)$.
Following \cite{PSbonn}, Section 2.1, and assuming the chaos property \eqref{pc}, Eq.\,\eqref{eq:SIR_Boltzmann} is obtained in the limit $N\to \infty, \e \to 0$ with $\e^{d-1}N=1$.

\subsection{Rigorous results}

We have derived formally the kinetic equations under proper scaling limits, presenting only the basic ideas. A rigorous approach is possible, based on existing literature. In the case of the stochastic system, one can apply martingale techniques as in \cite{W}, or the hierarchy of equations for the family of the marginals \cite {PWZ}, or coupling techniques \cite {GM97}. In the case of the mechanical model, one can resort to the validity techniques for the Boltzmann equation, leading to a short time result; see \cite{Lanford} and subsequent works \cite{IP89,Spohn,CIP,GSRT12E,PSS,PulvirSimonBG,De17}.

\bigskip
\bigskip
\bigskip

\ni {\bf \large Acknowledgments.} We are indebted to Gr\'{e}gory Miermont for enlightening comments inspiring this work.
We thank Nicola Cotugno for advice on bibliography, and Robert Patterson for sharing numerical simulations and for useful comments on the manuscript.

%

\end{document}